# Mid-infrared group-IV nanowire laser

Youngmin Kim[1,5†], Simone Assali[2†], Junyu Ge[3], Sebastian Koelling[2], Manlin Luo[1], Lu Luo[2], Hyo-Jun Joo[1], James Tan[1], Xuncheng Shi[1], Zoran Ikonic[4], Hong Li[1,3], Oussama Moutanabbir[2*] and Donguk Nam[1,6*]

[1]School of Electrical and Electronic Engineering, Nanyang Technological University, 50 Nanyang Avenue, Singapore 639798, Singapore

[2]Department of Engineering Physics, École Polytechnique de Montréal, C.P. 6079, Succ. Centre-Ville, Montréal, Québec H3C 3A7, Canada

[3]School of Mechanical and Aerospace Engineering, Nanyang Technological University, 50 Nanyang Avenue, Singapore 639798, Singapore

[4]School of Electronic and Electrical Engineering, University of Leeds, Leeds LS2 9JT, UK

[5]School of Materials Science and Engineering, Kookmin University, 77 Jeongneung-ro, Seongbuk-gu, Seoul 02707, Republic of Korea

[6]Department of Mechanical Engineering, Korea Advanced Institute of Science and Technology (KAIST), 291 Daehak-ro, Yuseong-gu, Daejeon 34141, Republic of Korea

†These authors contributed equally to this work.

*E-mail: oussama.moutanabbir@polymtl.ca, dwnam@kaist.ac.kr

## Abstract

Semiconductor nanowires have shown great potential for enabling ultra-compact lasers for integrated photonics platforms. Despite the impressive progress in developing nanowire lasers, their integration into Si photonics platforms remains challenging largely due to the use of III-V and II-VI semiconductors as gain media. These materials not only have high material costs, but also require inherently complex integration with Si-based fabrication processing, increasing overall costs and thereby limiting their large-scale adoption. Furthermore, these material-based nanowire lasers rarely emit above 2 μm, which is a technologically important wavelength regime for various applications in imaging and quantum sensing. Recently, group-IV nanowires, particularly direct bandgap GeSn nanowires capable of emitting above 2 μm, have emerged as promising cost-effective gain media for Si-compatible nanowire lasers, but there has been no successful demonstration of lasing from this seemingly promising nanowire platform. Herein, we report the experimental observation of lasing above 2 μm from a single bottom-up grown GeSn nanowire. By harnessing strain engineering and optimized cavity designs simultaneously, the single GeSn nanowire achieves an amplified material gain that can sufficiently overcome minimized optical losses, resulting in a single-mode lasing with an ultra-low threshold of ~5.3 kW cm$^{-2}$. Our finding paves the way for all-group IV mid-infrared photonic-integrated circuits with compact Si-compatible lasers for on-chip classical and quantum sensing and free-space communication.

## Introduction

Nanowire light sources play pivotal roles in a wide range of applications such as creating single-photon sources for quantum communication[1] and developing ultra-compact, efficient lasers[2]. In particular, nanowires have attracted a growing attention as a platform to facilitate lasing owing to nanowires' unique advantages such as intrinsic high quality and large optical confinement factor[3]. Over the past two decades, many research groups have made significant advances in nanowire lasers[4-10], including the first experimental observation of lasing in ZnO nanowires[4], electrical operation of CdS nanowire lasers[6], near-infrared lasing in GaAs nanowires[8], and mid-infrared lasing from InAs nanowires[9]. Despite the impressive progress in the research field, however, all bottom-up grown nanowire lasers reported to date have only utilized III-V and II-VI semiconductors, which are inherently complex to integrate with Si[11,12]. Moreover, it is rare for these nanowire lasers to emit wavelengths above 2 μm, a range that is critically important for Si-based quantum integrated circuits due to the reduced two-photon absorption and decreased Rayleigh scattering in Si at wavelengths greater than 2 μm[13].

Recently, nanowires made out of Si-compatible group-IV materials have garnered much attention as a cost-effective gain medium for lasing owing to new, ingenious techniques to achieve direct bandgap in group-IV materials and their ability to emit wavelengths above 2 μm. For example, Ge and SiGe nanowires can be transformed from indirect to direct bandgap materials by modifying the crystal structure from cubic to hexagonal[14]. While lasing from these direct bandgap hexagonal Ge and SiGe nanowires has been actively pursued recently[15], there has been no successful demonstration of lasing in these material systems yet, possibly owing to a low material gain and high optical losses.

GeSn alloys are another promising group-IV material for the realization of Si-compatible, ultra-compact nanowire lasers. GeSn becomes a direct bandgap material with Sn contents above ~8 at.%[12,16,17] thus enabling the demonstration of lasing in a thin film geometry[18-25]. There have been a number of attempts to achieve lasing in GeSn nanowires by comprehensively studying their optical properties[26-33], but the experimental demonstration of lasing remains elusive.

Here, we present the observation of mid-infrared lasing in a single bottom-up group-IV nanowire by harnessing strain engineering and optical cavity optimization. The optimized growth process utilizing ultra-thin 20 nm Ge cores as growth substrates enables the achievement of highly uniform Sn content and significantly reduced compressive strain in the GeSn shell, thereby securing ideal conditions for optical confinement within the nanowire. The residual intrinsic compressive strain in the GeSn shell is effectively mitigated by depositing a $SiO_X$ stressor layer, markedly improving the material gain of the gain medium. Concurrently, the cavity design is optimized through precise focused ion beam (FIB) milling, leading to a substantial reduction in optical losses within the single nanowire cavity. The improved material gain and reduced optical losses are substantiated by theoretical calculations employing the **k·p** method and numerical analyses, respectively. Remarkably, we observed a clear single mode lasing from strain-engineered and cavity-optimized single GeSn nanowires with an optical pumping threshold density of ~5.3 kW cm$^{-2}$, which is the lowest among all state-of-the-art strain-free GeSn lasers with a comparable Sn content at a similar operating temperature[19,21]. The validity of the observed lasing action is supported by obvious threshold behaviors exhibited in output power, as well as linewidth narrowing as a function of pump power. Our experimental observation of lasing in a single GeSn nanowire significantly bridges the gap between traditional compound semiconductor nanowire lasers and group IV nanowire lasers, paving the way towards realizing all-group-IV mid-infrared photonic-integrated circuits (PICs).

# Results

**Design of a strain-engineered and cavity-optimized single GeSn nanowire laser**

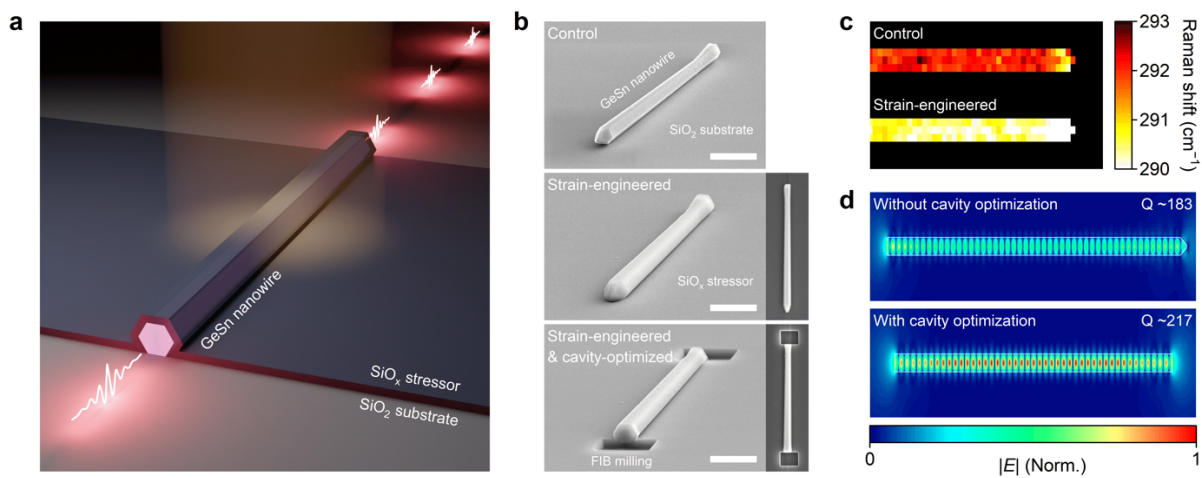

**Figure 1 | Design of a single GeSn nanowire laser with strain engineering and cavity optimization. a** Schematic illustration showing the laser emission in a single GeSn nanowire by harnessing strain engineering and cavity optimization. **b** Tilted-view SEM images of a single GeSn nanowire before (top) and after (middle) $SiO_X$ stressor deposition, followed by cavity optimization through FIB milling. Scale bar, 2 μm. Right insets: top-view SEM images of a single GeSn nanowire before (middle) and after (bottom) the cavity optimization. **c** Two-dimensional Raman map plotting the Raman peak position of a GeSn nanowire before (upper) and after (lower) strain engineering. A relatively uniform shift of ~2 cm$^{-1}$ in Raman peak positions confirms that the $SiO_X$ stressor technique can achieve high spatial uniformity in strain engineering. **d** Simulated top-view optical mode profiles of a single GeSn nanowire before (upper) and after (lower) cavity optimization. The Q-factor can be increased by employing cavity optimization due to the reduced optical losses at the facets.

Ge/Ge$_{0.90}$Sn$_{0.10}$ core/shell nanowires were synthesized using the vapor-liquid-solid (VLS) growth method, as discussed in more details in our previous reports[33] (also described in Methods for details on nanowire growth). The synthesized nanowires were then transferred to a SiO$_2$/Si substrate to achieve superior optical confinement[33]. Figure 1a presents a schematic illustration of a strain-engineered and cavity-optimized single GeSn nanowire laser. A stressed SiO$_X$ film is deposited onto the nanowire to mitigate the harmful intrinsic compressive strain in the GeSn shell. The facets of the nanowire are sharply etched to achieve the optimized cavity shape with reduced optical scattering at the facets.

Figure 1b shows tilted-view scanning electron micrograph (SEM) images capturing the nanowire in its control state (top), after strain engineering (middle), and after completion of both strain engineering and cavity optimization (bottom). The SEM image of a nanowire in its control state (top) confirms that a single GeSn nanowire is successfully transferred onto a SiO$_2$ layer, allowing detailed optical studies on a single nanowire. The SEM image of a strain-engineered nanowire (middle) shows the conformal deposition of the SiO$_x$ stressor layer achieved through plasma-enhanced chemical vapor deposition (PECVD), highlighting the excellence of uniform step coverage of our strain engineering technology (see Methods for more details on PECVD SiO$_x$ stressor). In the SEM image of a nanowire after both strain engineering and cavity optimization (bottom), it is shown that the nanowire facets are sharply etched by FIB milling, emphasizing the precise structural modification of the GeSn nanowire (see Methods for more details on FIB milling).

Figure 1c presents a two-dimensional (2D) Raman map that experimentally captures the Raman peak positions of a GeSn nanowire within a region of 7 × 3 μm$^2$, prior to and following the application of strain engineering (See Methods for more details on Raman spectroscopy). The 2D

Raman map reveals a spatially uniform Raman shift across the sample after the $SiO_x$ stressor deposition, confirming the high uniformity of the strain achieved through our strain engineering technique. The Raman peak position shifts by ~2 cm$^{-1}$, which indicates a compensation of compressive strain in GeSn via stressor deposition[34]. Owing to the lack of a reported Raman strain-shift coefficient for GeSn with a Sn content of ~10 at.%, we employ finite-element method (FEM) strain calculations to estimate strain compensation of ~0.3% along the <111> axial direction of the nanowire (see Supplementary Note 1 for more details on the estimation of the induced strain).

Figure 1d shows simulated top-view optical mode profiles of the nanowire, prior to (upper) and following (lower) cavity optimization. The nanowire without cavity optimization exhibits weak optical fields due to increased light scattering at the nanowire facets, whereas the cavity-optimized nanowire shows improved optical confinement owing to reduced scattering at its sharply etched facets. The simulated *Q*-factor is also increased by ~20%.

**Lasing emission from a single GeSn nanowire**

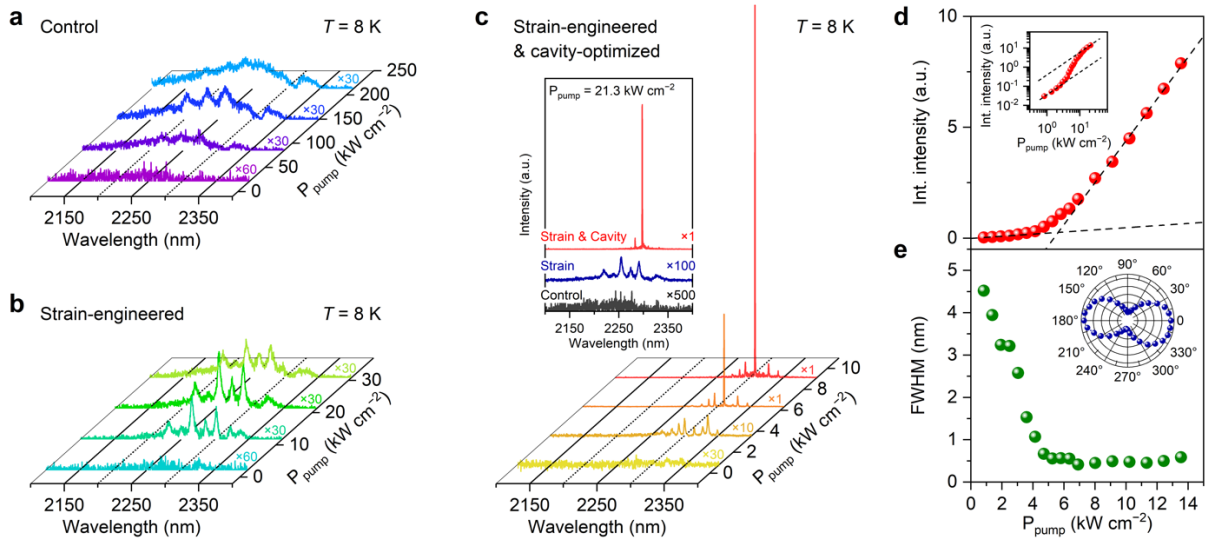

**Figure 2 | Lasing emission from a single GeSn nanowire at 8 K. a, b, c** Power-dependent photoluminescence spectra for control (**a**), strain-engineered (**b**), and strain-engineered & cavity-optimized (**c**) GeSn nanowires. The emission spectra from the control nanowire (**a**) show the onset, enhancement, and dissipation of the cavity resonances with an increase of the pump power from 21.3 to 212.8 kW cm$^{-2}$. For the strain-engineered nanowire (**b**), the emission spectra exhibit the emergence of sharper cavity resonances and their subsequent intensity reduction as the pump power increases from 2.1 to 30.9 kW cm$^{-2}$. In stark contrast, the emission from the strain-engineered & cavity-optimized nanowire (**c**) shows very sharp cavity modes that grow into lasing as the pump power is increased from 0.8 to 9.1 kW cm$^{-2}$. Inset: emission spectra of the three nanowires at a fixed pump power of 21.3 kW cm$^{-2}$, which clearly presents that strain engineering and cavity optimization enhance the emission properties of the GeSn nanowire. **d** L–L curve for a strain-engineered & cavity-optimized nanowire. Inset: corresponding double-logarithmic plot highlighting a nonlinear response to pumping power, represented by an S-shaped curve. **e** FWHM as a function of pump power for a strain-engineered & cavity-optimized nanowire, showing a

linewidth narrowing with increasing pump power. Inset: polarization dependence of the emission from the nanowire, indicating a highly polarized emission along the longitudinal axis of the nanowire.

To investigate the effects of strain engineering and cavity optimization on improving stimulated emission in GeSn nanowires, we conducted photoluminescence measurements on three samples shown in Fig. 1b: control, strain-engineered, and strain-engineered and cavity-optimized nanowires (See Methods for more details on photoluminescence spectroscopy). All measurements were performed at 8 K.

Figure 2a shows the emission spectra from a control nanowire at varying pump powers. At an initial pump power of 21.3 kW cm$^{-2}$, a broad spontaneous emission spectrum is observed. As the pump powers are increased to 85.1 and 148.9 kW cm$^{-2}$, three distinct cavity resonances emerge and become apparent within the wavelength range from 2224 to 2275 nm. However, these resonances dissipate at a further increased pump power of 212.8 kW cm$^{-2}$. This observed behavior of cavity resonances aligns with the findings from our recent study[33], which concluded that the cavity resonances do not grow into lasing due to an insufficient net gain, despite the successful achievement of cavity modes by enhanced optical confinement in the nanowire on an SiO$_2$ layer. In the study, we also discussed that one potential strategy for increasing net gain, and thereby achieving lasing, involves the application of strain engineering.

Figure 2b presents the emission spectra from a strain-engineered nanowire at various pump powers. It is evident that in this strain-engineered nanowire, cavity resonances emerge at a pump power of ~10 kW cm$^{-2}$, which is an order of magnitude lower compared to the required power to observe resonances in the control nanowire. While the resonance linewidth for the control nanowire at a pump power of 148.9 kW cm$^{-2}$ is ~13.5 nm, the linewidth for the strain-engineered nanowire at 11.7 kW cm$^{-2}$ is only ~4.6 nm. This narrow linewidth in the strained nanowire indicates an enhancement in the material gain[33], which will be explained in detail in the following theoretical modeling section. Also, the emission spectral range of the strain-engineered nanowire shifts to a longer wavelength in comparison to the control nanowire, which is indicative of a bandgap reduction in strain-engineered GeSn[16,20]. At a higher pump power of 30.9 kW cm$^{-2}$, however, the cavity modes in the strain-engineered nanowire become broader and weaker, which ultimately diminish at further increased pump powers. This result highlights that the net gain improvement through strain engineering alone is insufficient for achieving lasing.

Figure 2c shows the emission spectra of a strain-engineered and cavity-optimized nanowire at varied pump powers. At a low pump power of 3.6 kW cm$^{-2}$, very sharp cavity modes already appear. As the pump power is increased to 6.4 kW cm$^{-2}$, only one mode at 2310 nm grows rapidly in intensity by more than a magnitude while its linewidth becomes narrower, indicating the lasing action. At a higher pump power of 9.1 kW cm$^{-2}$, this mode continues to intensify, dominating all other neighboring cavity modes. The difference in emission characteristics among three different types of nanowires can be more clearly seen by comparing their emission spectra at one pump power of 21.3 kW cm$^{-2}$, as shown in the inset to Figure 2c. While the control nanowire exhibits weak spontaneous emission (black), the strain-engineered nanowire presents cavity resonances (blue) and the strain-engineered and cavity-optimized nanowire shows a significantly intensified lasing peak (red). This comparison confirms the effectiveness of strain engineering and cavity optimization in achieving lasing, which will be further explained in the following theoretical modeling section.

Lasing action in the strain-engineered and cavity-optimized nanowire is also quantitatively evidenced by the nonlinear behavior of integrated emission intensity and the reduction of linewidth as a function of pump power. Figure 2d presents the light-in-light-out (L–L) characteristics from the nanowire, revealing a superlinear enhancement in integrated intensity near the threshold pump power (~5.3 kW cm$^{-2}$). It is worth noting that the threshold power of ~5.3 kW cm$^{-2}$ is lower than those observed in state-of-the-art strain-relaxed GeSn lasers with similar Sn content achieved through GeSn thin film platforms[19,21]. This lowest threshold can be attributed to the minimal defect density in the nanowire that can be achieved through the bottom-up growth method[35]. An S-shaped curve in the corresponding double-logarithmic plot (inset) confirms a typical signature of lasing. A clear reduction in the full-width at half-maximum (FWHM) with increasing pump power is observed as shown in Figure 2d. The linewidth reduces from ~4.5 to 0.4 nm, further supporting the validity of the lasing action. Additionally, a highly polarized emission along the longitudinal axis of the nanowire is also observed as shown in the inset to Figure 2d. These highly polarized characteristics align with the findings from prior studies[36,37], which discussed that the nanowire structure has polarized characteristics along the longitudinal axis of the nanowire.
11

**Theoretical modeling for net gain in a strain-engineered and cavity-optimized GeSn nanowire**

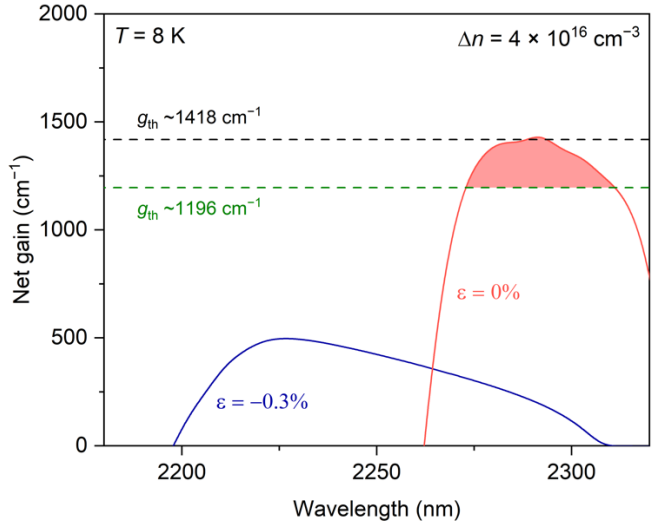

**Figure 3 | Theoretical modeling for net gain in a strain-engineered and cavity-optimized GeSn nanowire at 8 K.** Calculated net gains (solid lines) at an injection density of $4 \times 10^{16}$ cm$^{-3}$ for the nanowire under strains of –0.3% (blue) and 0% (red), which correspond to the strains in non-strain-engineered and strain-engineered nanowires, respectively. The threshold gains (dashed lines) before (black) and after (green) the cavity optimization are estimated to be ~1418 and ~1196 cm$^{-1}$, respectively.

To elucidate the role of strain engineering and cavity optimization in achieving lasing in the GeSn nanowire, we conducted theoretical modeling for the gain and loss dynamics in the GeSn under different strain levels (see Supplementary Note 2 for more details on the theoretical modeling). An injection density of $4 \times 10^{16}$ cm$^{-3}$, corresponding to a pump power density of 21.3 kW cm$^{-2}$, is utilized for the calculation. The material loss considers all possible loss mechanisms including free-carrier absorption (FCA) and inter-valence band absorption (IVBA). The net gain is determined by subtracting the loss from the gain for each strain.

Figure 3 shows the calculated net gain spectra (solid lines) at 8 K for $Ge_{0.90}Sn_{0.10}$ nanowires subjected to strains of −0.3% (blue) and 0% (red) corresponding to non-strain-engineered and strain-engineered nanowires, respectively. The estimated threshold gains of the nanowires with (black) and without (green) cavity optimization are also shown as horizontal dashed lines (see Supplementary Note 3 for more details on the threshold gain calculation). At the compressive strain of –0.3%, the net gain reaches its maximum intensity above ~2230 nm, corresponding to the peak of the broad spontaneous emission observed experimentally in the control nanowire (black curve of the inset to Figure 2c). As the compressive strain is compensated to yield a net strain of ~0%, GeSn becomes a more direct bandgap material due to increased directness, resulting in a considerable increase in the peak intensity of the net gain. This significant enhancement in net gain brings the net gain curve (red solid line) closer to the threshold gain for non-cavity-optimized nanowire (black dashed line), explaining the experimental observation of cavity modes from the strain-engineered nanowire (blue curve of the inset to Figure 2c). However, the transition from cavity modes to lasing does not occur, as the peak net gain (red solid line) does not dominantly surpass the theoretical threshold gain of 1418 $cm^{-1}$ for the non-cavity-optimized nanowire (black dashed line), highlighting the necessity of cavity optimization for achieving lasing. The cavity optimization leads to a reduction in the threshold gain to 1196 $cm^{-1}$ (green dashed line), thus enabling the net gain (red solid line) to exceed the threshold gain for lasing. This mechanism is corroborated by the experimental observation of lasing from the strain-engineered and cavity-optimized nanowire (red curve of the inset to Figure 2c). The correspondence of the measured lasing peak at 2298 nm to the theoretical net gain peak at 2292 nm further validates the consistency between our experimental results and the theoretical modeling.

**Temperature-dependent emission characteristics**

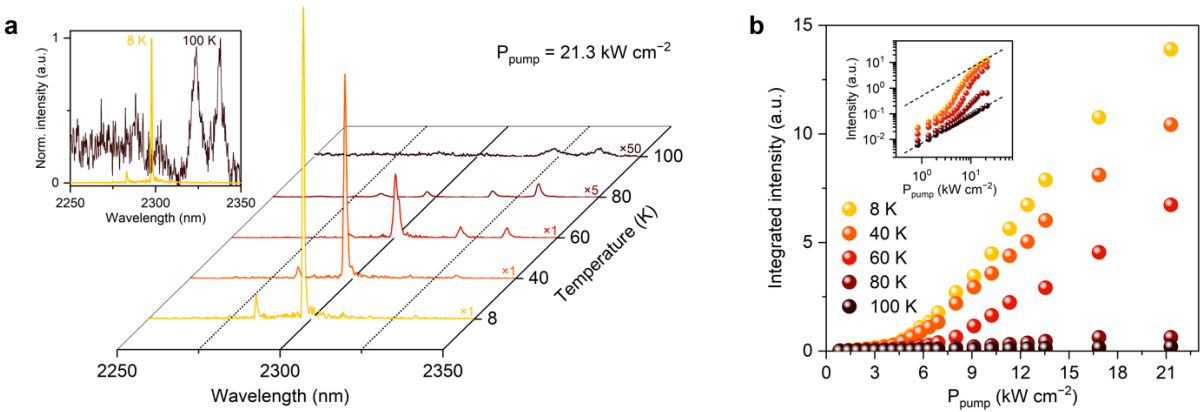

**Figure 4 | Temperature-dependent emission characteristics. a** Emission spectra for a strain-engineered and cavity-optimized GeSn nanowire at temperatures of 8 K (yellow), 40 K (orange), 60 K (red), 80 K (dark red) and 100 K (dark brown), all under a constant pump power of 21.3 kW cm$^{-2}$. Inset: normalized emission spectra at 8 K (yellow) and 100 K (dark brown). **b** L–L curves measured at temperatures of 8 K (yellow), 40 K (orange), 60 K (red), 80 K (dark red) and 100 K (dark brown). Inset: corresponding double-logarithmic plots for varied temperatures.

To explore the temperature-dependent lasing characteristics of the strain-engineered and cavity-optimized GeSn nanowire, we performed temperature-dependent photoluminescence measurements. Figure 4a presents emission spectra observed at temperatures of 8 K (yellow), 40 K (orange), 60 K (red), 80 K (dark red) and 100 K (dark brown) with a fixed pump power of 21.3 kW cm$^{-2}$. As the temperature is elevated from 8 K to 100 K, a notable reduction in emission intensity is observed. This attenuation in emission intensity is primarily due to an enhanced loss mechanism associated with an increased carrier scattering from Γ to L valleys[18] and a heightened rate of nonradiative recombination[19], both of which are more severe at higher temperatures. The inset to Figure 4a illustrates the normalized emission spectra at 8 K (yellow) and 100 K (dark brown), contrasting a sharp lasing mode at 8 K and broadened cavity modes at 100 K due to the increased loss at a higher temperature mentioned earlier.

Figure 4b shows the L–L characteristics of the nanowire at various temperatures of 8 K (yellow), 40 K (orange), 60 K (red), 80 K (dark red) and 100 K (dark brown). The corresponding double-logarithmic plots are also presented as an inset. At 8 K, 40 K and 60 K, the L–L curves exhibit a distinct superlinear behavior and an S-shaped profile near the threshold. As the temperature increases to 80 K, while the emission retains its superlinear nature, the S-shaped profile becomes less pronounced. As the temperature is further increased to 100 K, the L–L curve only shows a linear behavior, indicating the absence of lasing action. The operational temperature of our nanowire platform is similar to that of strain-free GeSn lasers with similar Sn content in GeSn film platforms[19,21]. Despite the lower threshold in our nanowire, the operational temperature is comparable, primarily because Sn content is a dominant factor in determining the maximum operating temperature[19].

**Discussion**

In summary, we presented the first observation of mid-infrared lasing in a single group-IV nanowire. By leveraging strain engineering and cavity optimization, we achieve a distinct single-mode lasing with a threshold power density of ~5.3 kW cm$^{-2}$, which is the lowest among all reported GeSn lasers with comparable Sn contents[19,21]. The lasing action is evidence by clear threshold characteristics including the superlinear increase in integrated intensity and a narrowing of linewidth as a function of pump power. This achievement of lasing relies on the growth of nanowires with enhanced strain relaxation and uniform Sn content leveraging 20 nm-diameter Ge nanowires as compliant substrates in addition to the compensation of detrimental residual compressive strain via the strain engineering approach and the reduction of optical field scattering within the cavity through a refined cavity optimization technique. The deposition of an SiO$_x$ stressor layer effectively compensates the intrinsic compressive strain in the GeSn shell, significantly enhancing the net gain of the GeSn gain medium. The refinement of nanowire facets through FIB milling reduces optical field scattering at facets, markedly lowering the optical loss in the cavity. The enhanced net gain and the reduced optical loss are confirmed by comprehensive theoretical calculations based on the **k·p** method and numerical analyses, respectively, explaining the mechanism behind achieving lasing in a single GeSn nanowire. Our experimental demonstration of lasing at wavelengths above 2μm in a single group-IV nanowire sets the stage for the development of monolithic mid-infrared PICs long sought-after for on-chip classical and quantum sensing and free-space communication.

## Methods

### Nanowire growth

The Ge/Ge$_{0.90}$Sn$_{0.10}$ core/shell nanowires were synthesized in a chemical vapor deposition (CVD) reactor using germane (GeH$_4$) and tin tetrachloride (SnCl$_4$) as precursor gases. Before the growth process, the Ge(111) substrate was cleaned with a 2% HF solution. Subsequently, 20 nm Au colloids were deposited on the wafer, dried with a nitrogen flow, and then placed in the CVD reactor. The vapor-liquid-solid (VLS) method was used to grow 20 nm diameter Ge core nanowires for 20 minutes at 340 °C, followed by the growth of the Ge$_{0.90}$Sn$_{0.10}$ shell for 150 minutes at 310 °C (with a Ge/Sn gas phase ratio of 2200).

### Plasma-enhanced chemical vapor deposition SiO$_x$ stressor

To induce tensile strain within the GeSn shell, compressively stressed SiO$_x$ was deposited on the nanowire using plasma-enhanced chemical vapor deposition (PECVD). The deposition process was conducted at a pressure of 270 mTorr, utilizing a gas mixture of SiH$_4$ and N$_2$O with flow rates of 40 sccm and 20 sccm, respectively. The deposition power was maintained at 100 W. Wafer curvature assessments on test wafers, employing the same deposition technique, revealed an intrinsic compressive stress of –470 MPa within the SiO$_x$ film, which had a thickness of 207 nm.

### Focused ion beam milling

The structural profile of the nanowire cavity was meticulously refined using focused ion beam (FIB) milling, which employed high-energy gallium ions. The milling process was precisely conducted with a current of 5 pA, a dwell time of 5 µs, and a voltage of 30 kV. The targeted milling area for each facet was precisely set to 1.5 x 1.5 µm², ensuring that both facets were uniformly and finely milled.

**Raman spectroscopy**

To examine strain relaxation in the GeSn nanowire before and after the application of strain engineering technique, spatially resolved Raman spectroscopy was performed. A laser pump with a wavelength of 532 nm was precisely focused onto the surface of the nanowire using a 100× objective lens. The sample was placed on a piezo-electric stage, which facilitated in-plane step displacement adjustments of 150 nm. Subsequently, the Raman spectrum was captured at each position, and the data were processed to generate µ-Raman maps of the nanowire surface. Throughout the measurement process, careful attention was given to maintaining a low laser power to prevent any potential heating effects.

**Photoluminescence spectroscopy**

The samples were securely mounted within a closed-cycle helium cryostat (Cryostation s50, Montana Instrument) to facilitate investigations at low and varied temperatures. A 1550 nm laser, which has a 5 ns pulse width and a 1 MHz repetition rate, was focused onto the samples through a 15× reflective objective lens. The same lens was employed to collect the emitted signals from the samples. Subsequently, the collected signals were directed into a Fourier Transform Infrared (FTIR) interferometer and detected by an extended InGaAs detector.

# Acknowledgements

This research conducted in Singapore is supported by the National Research Foundation, Singapore and A*STAR under its Quantum Engineering Programme (NRF2022-QEP2-02- P13). The authors would like to acknowledge and thank the Nanyang NanoFabrication Centre (N2FC). The work carried out in Montreal is supported by NSERC Canada, Canada Research Chairs, Canada Foundation for Innovation, Mitacs, PRIMA Québec, Defence Canada (Innovation for Defence Excellence and Security, IDEaS), the European Union's Horizon Europe research and innovation program under grant agreement No. 101070700 (MIRAQLS), the US Army Research Office Grant No. W911NF-22-1-0277, and the Air Force Office of Scientific and Research Grant No. FA9550-23-1-0763.

## Author contributions

Y.K., S.A., O.M., and D.N. conceived the initial idea of the project. Under the guidance of O.M., S.A. grew the nanowires. S.A., S.K., and L.L. carried out material characterization and analyses of the as-grown nanowires. Under the guidance of D.N., Y.K., H.-J. J., J.T., and X.S. conducted plasma-enhanced chemical vapor deposition. Under the guidance of H.L. and D.N., J.G., Y.K, and M.L. performed focused ion beam milling. Y.K. carried out photoluminescence measurements, Raman measurements, strain simulations, optical simulations. Z.I. performed theoretical net gain calculation and modeling. Under the guidance of O.M. and D.N., Y.K. and S.A. conducted data analyses. Y.K., S.A., O.M., and D.N. wrote the manuscript. All authors revised the manuscript.

## Competing interests

The authors declare no competing interests.